\documentclass[nofootinbib,prd,reprint,preprintnumbers,twocolumn]{revtex4}
\usepackage{amsmath,amssymb,bm,mathrsfs}
\usepackage{graphicx,color}
\usepackage{amsfonts}
\usepackage[export]{adjustbox}
\usepackage{mathtools}
\usepackage[sort&compress]{natbib} 
\usepackage{srcltx}
\usepackage{dsfont}
\usepackage{epsfig}
\usepackage{bbold}
\usepackage{bbm}  
\usepackage{rotating}
\usepackage[dvipsnames]{xcolor}
\usepackage{subfig}
\usepackage{xfrac}
\usepackage{multirow}
\usepackage{booktabs}
\usepackage[inline]{enumitem} 
\usepackage[colorlinks=true
,urlcolor=blue
,anchorcolor=blue
,citecolor=blue
,filecolor=blue
,linkcolor=red
,menucolor=blue
,linktocpage=true
,pdfproducer=medialab
,pdfa=true
]{hyperref}
\usepackage{cleveref}
\usepackage{setspace}

\AtBeginDocument{%
  \heavyrulewidth=.08em
  \lightrulewidth=.05em
  \cmidrulewidth=.03em
  \belowrulesep=.65ex
  \belowbottomsep=0pt
  \aboverulesep=.4ex
  \abovetopsep=0pt
  \cmidrulesep=\doublerulesep
  \cmidrulekern=.5em
  \defaultaddspace=.5em
} 

\usepackage{tikz}

\definecolor{forestgreen}{RGB}{34,139,34}
\definecolor{nanocolor}{RGB}{15.5, 59.5, 90.}
\definecolor{eptacolor}{RGB}{0., 95.625, 191.25}
\definecolor{pptacolor}{RGB}{0., 162.563, 191.25}
\definecolor{SMBHcolor}{RGB}{255., 127., 14.}
\definecolor{Ourcolor}{RGB}{142.667, 26., 26.6667}

\newcounter{qnumber}

\begin{document}

\title{Searching For Stochastic Gravitational Waves Below a Nanohertz}

\author{William DeRocco}

\affiliation{Department of Physics, University of California Santa Cruz, 1156 High St., Santa Cruz, CA 95064, USA}
\affiliation{Santa Cruz Institute for Particle Physics, 1156 High St., Santa Cruz, CA 95064, USA}

\author{Jeff A. Dror}

\affiliation{Department of Physics, University of Florida, Gainesville, FL 32611, USA}
\affiliation{Department of Physics, University of California Santa Cruz, 1156 High St., Santa Cruz, CA 95064, USA}
\affiliation{Santa Cruz Institute for Particle Physics, 1156 High St., Santa Cruz, CA 95064, USA}

\begin{abstract}
The stochastic gravitational-wave background is imprinted on the times of arrival of radio pulses from millisecond pulsars. Traditional pulsar timing analyses fit a timing model to each pulsar and search the residuals of the fit for a stationary time correlation. This method breaks down at gravitational-wave frequencies below the inverse observation time of the array; therefore, existing analyses restrict their searches to frequencies above 1~nHz. An effective method to overcome this challenge is to study the correlation of secular drifts of parameters in the pulsar timing model itself. In this paper, we show that timing model correlations are sensitive to sub-nanohertz stochastic gravitational waves and perform a search using existing measurements of pulsar spin-decelerations and pulsar binary orbital decay rates. We do not observe a signal at our present sensitivity, constraining the stochastic gravitational-wave relic energy density to $\Omega_\text{GW} ( f )  < 3.8 \times 10 ^{ - 9} $ at 450~pHz with sensitivity which scales as the frequency squared until approximately 10~pHz. We place additional limits on the amplitude of a power-law spectrum of $A_\star \lesssim 1.8\times10^{-14}$  for a reference frequency of $f_* = 1~{\rm year} ^{-1} $ and the spectral index expected from supermassive black hole binaries, $\gamma = 13/3$. If detection of a supermassive black hole binary signal above 1~nHz is confirmed, this search method will serve as a critical complementary probe of the dynamics of galaxy evolution.
 \end{abstract}

\maketitle
\section{Introduction}
The detection of gravitational waves (GWs) across the frequency spectrum is one of the most pressing goals for fundamental physics in the twenty-first century. Using a combination of cosmic microwave background~\cite{Planck:2018jri}, pulsar timing~\cite{NANOGrav:2020bcs,Goncharov:2021oub,Chen:2021rqp,Antoniadis:2022pcn,Tarafdar:2022toa}, and laser interferometry~\cite{KAGRA:2021kbb} data, existing searches cover a huge range of frequencies from as low as $ 10 ^{ - 18}~{\rm Hz} $ to as high as a $ {\rm kHz} $, with proposals to explore even higher frequencies~\cite{Aggarwal:2020olq,Domcke:2022rgu,Berlin:2021txa}. Nevertheless, our frequency coverage has two prominent gaps: one at $ 10 ^{ - 16}~ {\rm Hz}-{\rm nHz} $ and one at $ 100~{\rm nHz} - 10~ {\rm Hz} $. While there has been a significant effort to cover the latter gap using a combination of ground-based and space-based techniques~\cite{2017arXiv170200786A,Dimopoulos:2007cj,Maggiore:2019uih,Pyne:1995iy,Book:2010pf,PhysRevLett.119.261102,Klioner:2017asb,Wang:2020pmf,Fedderke:2022kxq,Fedderke:2021kuy,Blas:2021mqw,Blas:2021mpc}, the sub-nHz gap is still largely unexplored. Below 1~nHz, the frequency of gravitational waves is below the current inverse observation times of experiments ($\sim 30$ yr), which we refer to as the {\em ultralow}-frequency regime. Such GWs do not appear as periodic variations in data but rather as secular drifts in experimental observables. As we demonstrated in Ref.~\cite{DeRocco:2022irl}, these drifts are detectable in the fit parameters of pulsar timing models, providing a new means to probe the frequency spectrum in the sub-nHz gap.\footnote{Note that other proposals to use pulsar timing models to probe this regime appear in the literature~\cite{10.1093/mnras/203.4.945,Kopeikin:1997rj,1999MNRAS.305..563K,2003AstL...29..241P,Kopeikin:2004hw,2009MNRAS.398.1932P,2018MNRAS.478.1670Y,2019MNRAS.489.3547K,Kumamoto:2020nas,Kikunaga:2021wwu}, however they were substantially less sensitive then the methodology presented in Ref.~\cite{DeRocco:2022irl}. See also Refs.~\cite{Gwinn:1996gv,Book:2010pf,Darling:2018hmc,Jaraba:2023djs} for discussions of using astrometric lensing as a complementary probe of this frequency range.} 

Gravitational waves in the ultralow-frequency regime are strongly motivated by the existence of supermassive black hole (SMBH) binaries~\cite{1980Natur.287..307B}, which are expected to give rise to a signal within reach of current pulsar timing array (PTA) analyses. Recent measurements of a signal at frequencies above 1~nHz by NANOGrav~\cite{NANOGrav:2020bcs}, EPTA~\cite{Chen:2021rqp}, and PPTA~\cite{Goncharov:2021oub} hint at this source, and if the spatial correlations of this signal are confirmed, it will mark the first detection of gravitational waves in the nanohertz regime. If the discovery is confirmed, it implies a likely signal waiting to be found below 1~nHz. Various possible cosmic sources can also give rise to ultralow-frequency gravitational waves, such as cosmic strings~\cite{Chang:2019mza,ghoshal2023primordial}, bubble collisions~\cite{Freese:2023fcr}, and a turbulent QCD phase transition~\cite{Neronov:2020qrl,Brandenburg:2021tmp} (see also, Ref.~\cite{Moore:2021ibq} and references therein), providing further motivation to probe this frequency range.

In Ref.~\cite{DeRocco:2022irl}, we studied GW detection using measurements of the pulsar timing model, focusing on two key parameters: the second time derivative of the pulsar period ($ \ddot{ P} $) and the first derivative of the binary period ($ \dot{ P} _b $). By searching for correlations in the parameter values for different pulsars in the sky, we demonstrated the method could detect GWs with comparable amplitude to traditional techniques at 1 nHz and is robust against astrophysical uncertainties. We then applied this method to search for localized, continuous gravitational-wave sources. This paper extends this analysis to search for a stochastic gravitational-wave background (SGWB), the signal induced by a sum of incoherent GW sources.

Detecting an SGWB in the ultralow-frequency regime requires revisiting how stochastic GWs are imprinted on the timing of pulsars. To this end, we derive expressions for the correlation between timing model parameters. We exploit this result to search for the SGWB in existing data sets of pulsar timing parameters. While we do not detect a signal in the data used for our analyses, we find sensitivity comparable to that of existing PTA analyses near 1~nHz and comment on future detection prospects.

\begin{figure*}[]
\begin{center} 
\includegraphics[width=17cm]{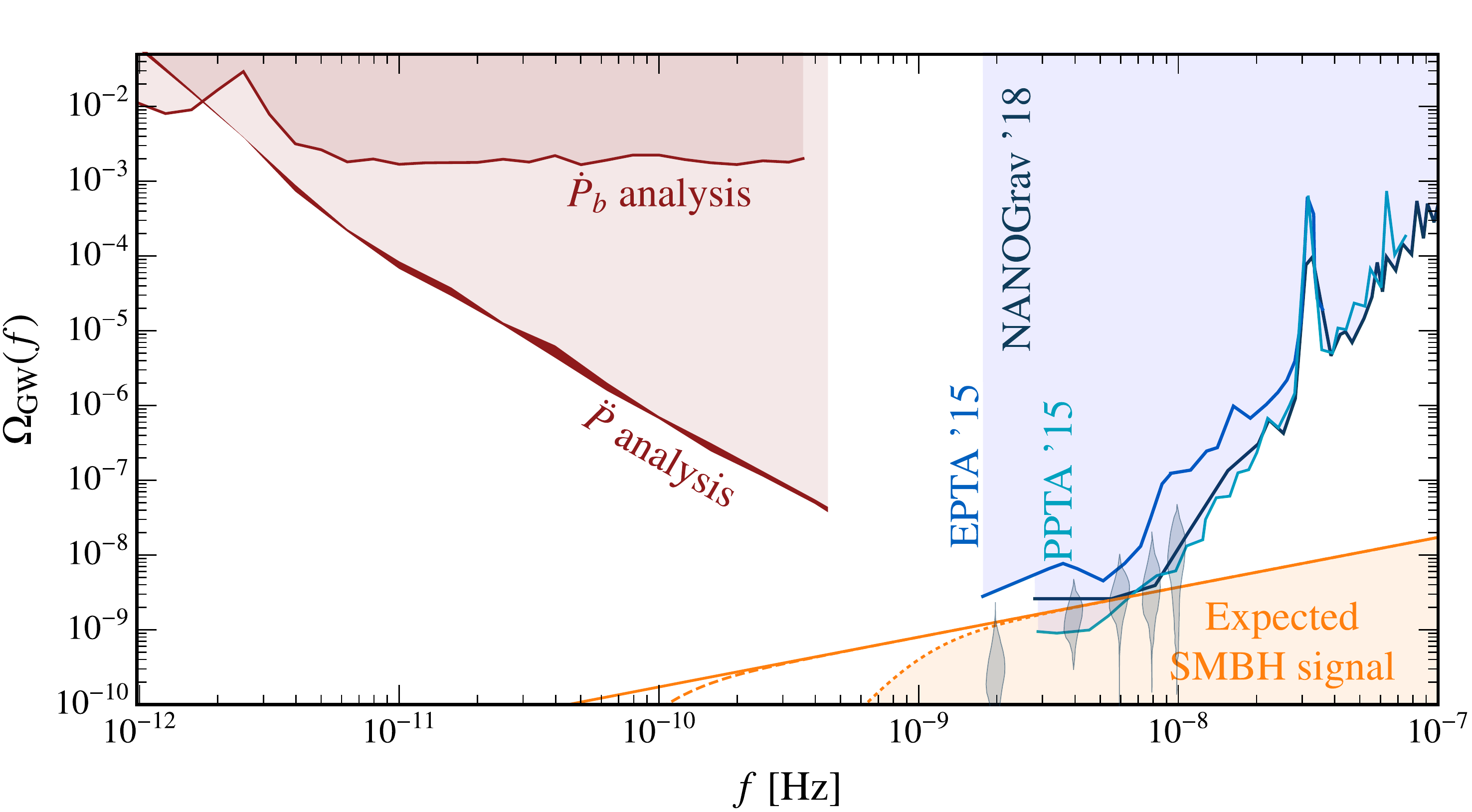} 
\end{center}
\caption{The sensitivity of pulsar timing parameters to the ultralow-frequency SGWB ({\bf \color{Ourcolor} red}) as a function of frequency. We show analysis results using existing measurements of the second derivative of the pulsar period ($ \ddot{ P} $) and the binary orbital period derivative ($ \dot{ P} _b $). For the $ \ddot{ P} $ measurement, ultralow-frequency red noise can be important, and we use the width of the line to encompass the possible influence of this noise (see text for details). We also show results of free-spectrum analyses from NANOGrav ({\bf\color{nanocolor}dark blue})~\cite{NANOGRAV:2018hou}, EPTA ({\bf\color{eptacolor}blue})~\cite{Lentati:2015qwp}, and PPTA ({\bf \color{pptacolor}light blue})~\cite{Shannon:2015ect}, which while not formally equivalent to the quantity we limit, provide a qualitative comparison. We also present the expected signal from supermassive black hole mergers, assuming the tentative signal above 1~nHz persists ({\bf\color{SMBHcolor}yellow}). The lower frequency part of the SMBH spectrum is taken to be driven by stellar scattering, and we show a few possible bending frequencies~\cite{Sampson:2015ada}. Finally, we show the best-fit region of the gravitational wave amplitude of the recent detection by the NANOGrav collaboration in {{\bf \color{gray} gray}}~\cite{NANOGrav:2023gor}. For further discussion of this figure, see Sec.~\ref{sec:mono}.} 
\label{fig:sensitivity}
\end{figure*}

The paper is organized as follows. In Sec.~\ref{sec:review}, we review the results of Ref.~\cite{DeRocco:2022irl}, presenting two observables within the pulsar timing model that are particularly sensitive to ultralow-frequency gravitational waves. In Sec.~\ref{sec:ultralow}, we calculate the impact of a stochastic background on these observables and discuss the behavior of the signal. In Sec.~\ref{sec:methods}, we present the data sets and statistical methodology we employ to search for a signal. In Sec.~\ref{sec:applications}, we apply the search to three possible gravitational-wave spectra and discuss the results in Sec.~\ref{sec:discussion}.

\section{Gravitational-Wave Detection Using Parameter Drifts}
\label{sec:review}

Pulsar timing arrays measure the pulse arrival times of the order of a hundred stable millisecond pulsars over decades-long timescales. The times of arrival (TOAs) are measured with high accuracy and are sensitive to small deviations induced by gravitational waves. In a conventional PTA analysis, a timing model is first fit to the arrival times to model deterministic trends in the TOAs, due to, e.g., quadratic spin-down of the pulsar. Once the timing model parameters are fit, the model prediction for the TOAs is subtracted from the data to produce the \textit{timing residuals}. GWs with frequencies above the inverse observation time of the array induce correlations in these residuals that serve as the target of conventional analyses. In contrast, GWs with frequencies below the inverse observation time of the array appear as deterministic trends in the data. As a result, if the timing model is sufficiently extensive, the fitting procedure removes ultralow-frequency GWs from the residuals. Nevertheless, the best-fit values of the fit parameters themselves remain sensitive to GWs.

The set of timing model parameters used to describe a particular pulsar varies but always includes the pulsar period ($ P $), its first-time derivative ($ \dot{ P} $), and potentially its second-time derivative ($ \ddot{ P} $). The parameters for pulsars in a binary system include the binary period ($ P _b $) and its time derivative ($ \dot{ P } _b  $). In Ref.~\cite{DeRocco:2022irl}, we advocated for studying two timing model parameters sensitive to the presence of GWs to search for ultralow-frequency GWs: $ \dot{ P} _b $ and $ \ddot{ P} $. For these parameters, the best-fit values (denoted by subscript ``obs'' for ``observed'') comprise of a sum of distinct, independent physical effects. 

The change in the binary period is given by, 
\begin{equation} 
\frac{ \dot{ P}_{b,{\rm obs}}}{  P_b  }   = \frac{ \dot{ P } _{b, {\rm int}}}{P_b} -  \frac{ v _{ \perp } ^2 }{ d  }  - a _{ \text{MW}} - a _{ \text{GW}}\,, \label{eq:Pbd}
\end{equation}
where the left-hand side is the observed value. The first term on the right-hand side is the intrinsic decrease in the binary period from gravitational-wave emission,\footnote{Note that these are \textit{not} the ultralow-frequency GWs of interest, but rather the gravitational radiation generated by binary motion.} which can be calculated with independent measurements of the companion mass and orbital parameters. The second is a kinematic effect due to proper motion on the sky ($ v _\perp $) of a pulsar a distance $ d  $ away. These can be measured independently using timing data at higher frequencies or Very Long Baseline Interferometry techniques. The third term is due to acceleration induced by the Milky Way potential. While typically negligible, this contribution can be estimated from models of the galactic mass distribution. The final term is the target of our analysis, the acceleration induced by ultralow-frequency GWs. Subtracting the first three contributions from the observed value gives a measurement of $a_\text{GW}$.

The case of $\ddot{P}$ is considerably simpler; all non-GW contributions can be estimated to be far below current experimental error bounds (see App.~B of Ref.~\cite{DeRocco:2022irl}). Therefore, the only term contributing to a non-zero $\ddot{P}_\text{obs}/P$ is due to the jerk induced by GWs:
\begin{equation}
\frac{ \ddot{ P }_{\text{obs}}}{P} =  j _{\text{GW}}\,.   \label{eq:Pdd}
\end{equation}

Gravitational waves induce a strain, which we denote as $h_{ij} ^{ TT}(\mathbf{x},t)$ in the traceless-transverse gauge, which is observable as an apparent line-of-sight velocity between the solar system barycenter (SSB) and a pulsar.\footnote{While $ v _{ {\rm GW}} $ acts like a velocity when considering a single line of sight, the quadrupolar structure of gravitational waves induces correlations between different lines of sight that are not the dipolar correlations that would be induced by simple motion along an axis.}  This velocity is given as an integral over the line of sight (see, e.g., App.~A of Ref.~\cite{DeRocco:2022irl}),
\begin{align} 
v _{ {\rm GW}} ^{ ( a ) }( t ) &  = \frac{1}{2} \hat{{\bf n}}_a^i \hat{{\bf n}}_a^j \int_{t - d _a  }^{t } dt' \left[\frac{\partial}{\partial t'} h_{ij}^{\text{TT}}(t', \mathbf{x})\right]_{\mathbf{x}=\mathbf{x}_0 ^{ (a) }(t')}\,,
\label{eq:vEP}
\end{align} 
where $\hat{{\bf n}} _a $ denotes the direction vector pointing from the SSB to the pulsar, $d_a$ is the SSB-pulsar distance, $ {\bf x _0} ^{ ( a ) }( t' ) \equiv ( t  - t ' )  \hat{{\bf n}} _a $, and we introduce an index $ a $ to specify values for pulsar $ a $. Note that the derivative acts only on the first argument of $h_{ij} ^{ {\rm TT}}$. One can calculate $a_\text{GW} ^{ ( a ) }$ and $j_\text{GW} ^{ ( a ) }$ by taking temporal derivatives of Eq.~\eqref{eq:vEP}.

\section{Stochastic Signals at Ultralow Frequencies}
\label{sec:ultralow}
The stochastic gravitational-wave background has been studied extensively at higher frequencies (see, e.g., Ref.~\cite{SGWBreview} for a review). However, the signal changes dramatically in regimes at which the frequency falls below the inverse observation time. We carry out the analysis assuming the timing model fit incorporates $ \ddot{ P} $ and $ \dot{ P } _b $ as fit parameters. If this is the case, ultralow-frequency signals have been removed from the residuals by the fitting procedure and now reside in the fit parameters themselves.~\footnote{If $ \ddot{ P} $ or $ \dot{ P} _b $ is \textit{not} included in the fit, the residuals can, in principle, still be used to search for ultralow frequency GWs. In this case, the residual two-point correlator is not stationary and looks quite different from its traditional form. We derive the residual correlator and show how it reduces to the stationary form for frequencies above $ 1 /T $ in App.~\ref{app:residuals}. We note that it has been suggested in the literature that one could sample the stationary correlator at frequencies below $ 1/T $ (see, e.g., Ref.~\cite{10.1093/mnras/stt2122}). This will not capture the full influence of the gravitational wave signal.} 

The timing model parameters in Eqs.~\eqref{eq:Pbd} and \eqref{eq:Pdd} are sensitive to the secular motion of the SSB-pulsar system. The corresponding acceleration and jerk induced by ultralow-frequency GWs are given by derivatives of Eq.~\eqref{eq:vEP} if the GW is well approximated by its Taylor expansion around $ t = 0 $. We take this requirement to correspond to restricting the frequency to be less than $f _{ T} \equiv 1/4T _{ {\rm max}} $, where $ T _{ {\rm max}} $ is the maximum observation time of a pulsar in a data set. 

To carry out a search for an SGWB, we must understand how $ a _{ {\rm GW}} ^{ ( a ) } $ and $ j _{ {\rm GW}} ^{ ( a ) } $ are distributed. Since they arise as time-derivatives of $ v _{ {\rm GW}} ^{ ( a ) } $, they are functionals of the gravitational field; since the SGWB $ h _{ij} $ behaves as a Gaussian random field, they obey a Gaussian distribution of mean zero with correlations entirely specified by their covariance matrix. 

For an SGWB, $a_\text{GW}^{(a)}$ and $j_\text{GW}^{(a)}$ are intrinsically random variables. For gravitational waves with frequencies well above $ f _T $, their values (along with higher derivatives) are sampled repeatedly within the TOAs, and their variance is approximately given by its ergodic average (even in the case of data from a single pulsar). In contrast, in the ultralow-frequency regime, they take on approximately constant values for the observation time of the experiment; as such, their values exhibit significant cosmic variance. Owing to this, instead of using temporal correlations to estimate their variance, one must study pulsar-pulsar correlations. 

To study the acceleration and jerk distribution, we begin with the Fourier transform of the gravitational wave field split into the $ + $ and $ \times $ polarization,
\begin{equation} 
h _{ij} ^{ {\rm TT}} = \sum _{ A = + , \times } \int _{ - \infty } ^{\infty} df \int d ^2 \hat{{\bf n}}\, \tilde{h} _A ( f , \hat{{\bf n}} ) e_{ij}^A(\hat{{\bf n}})e ^{ - 2\pi i f ( t - \hat{{\bf n}} \cdot {\mathbf{x}} ) }\,,
\label{eq:hTT}
\end{equation}
where $e_{ij}^A(\hat{{\bf n}})$ is the $+$ or $ \times$ polarization tensor. Inserting Eq.~\eqref{eq:hTT} into Eq.~\eqref{eq:vEP} gives an expression for the apparent relative velocity between the SSB and the pulsar,
\begin{align} 
\label{eq:vgw}
v _{ {\rm GW}} ^{ ( a ) } & = \sum _{ A = + , \times } \int _{ - \infty }^{\infty} df \int d ^2 \hat{{\bf n} }\, \tilde{h} _A ( f ,  \hat{{\bf n}} ) F _a ^A (  \hat{{\bf n}} ) e ^{ - 2\pi i f t } \notag \\ 
& \times \left[ 1 - e ^{ 2\pi i f d _a ( 1 + \hat{{\bf n}} \cdot \hat{{\bf n}} _a ) } \right] \, ,
\end{align} 
where the $F _a ( \hat{{\bf n}} )$ are the ``pattern functions'' associated with a gravitational plane wave,
\begin{equation} 
F  _{ A } ( \hat{{\bf n}} )  \equiv \frac{ \hat{{\bf n} } _a ^i \hat{{\bf n} } _a ^j \hat{e} _{ij} ^A ( \hat{{\bf n}} ) }{ 2 ( 1 + \hat{{\bf n}} \cdot { \hat{\bf n}} _a ) }\,.
\end{equation}
The power spectrum of the gravitational field is defined via its correlator,
\begin{equation} 
\big\langle \tilde{h} _A ^\ast ( f  , \hat{{\bf n}} ) \tilde{h} _{A '} ( f ' , \hat{{\bf n}} ' ) \big\rangle = \delta ( f - f ' ) \frac{ \delta  ^2 ( \hat{{\bf n}} , \hat{{\bf n}} ' ) }{ 4\pi } \delta _{ A A'} \frac{1}{2} S _h ( f ) \,,
\end{equation} 
and upon substitution into Eq.~\eqref{eq:vgw}, this yields the velocity two-point function,
\begin{equation} 
 \big\langle v _{{\rm GW}} ^{ ( a ) }  v _{ {\rm GW}} ^{ ( b ) } \big\rangle = \frac{1}{2} \int _{ - \infty   } ^{\infty  } S _h ( f ) C ( \theta _{ ab }, f ) \, df \,,
\label{eq:vGWvGW}
\end{equation} 
 where $ \theta _{ ab } $ is the angle between the two pulsars,
 \begin{equation} 
C ( \theta _{ ab}, f ) \equiv \int \frac{ d ^2 \hat{{\bf n}} }{ 4\pi } {\cal K} _{ a b} ( f , \hat{{\bf n}} ) \sum _{ A } F _a ^A ( \hat{{\bf n}} ) F _b ^{ A} ( \hat{{\bf n}} ) \,,
\end{equation} 
and
\begin{equation} 
\label{eq:K}
{\cal K} _{ a b }( f , \hat{{\bf n}} ) \equiv \left[ 1 - e ^{ - 2\pi i f d _a ( 1 + \hat{{\bf n}} \cdot \hat{{\bf n}} _a ) } \right] \left[ 1 - e ^{ 2\pi i f d _b ( 1 + \hat{{\bf n}} \cdot \hat{{\bf n}} _b ) } \right]\,.
\end{equation}
The result in Eq.~\eqref{eq:vGWvGW}, is well known in the literature (see, e.g., Refs.~\cite{1983ApJ...265L..39H,PhysRevD.90.062011,Taylor:2021yjx}). This correlator can be extended to acceleration and jerk by taking time derivatives of Eq.~\eqref{eq:vgw} prior to averaging. Carrying out the calculation and introducing a high-frequency cut-off at $ f _T $ results in
\begin{equation} 
  \big\langle a _{ {\rm GW}} ^{ ( a ) }  a _{ {\rm GW}} ^{ (b ) }  \big\rangle = \int _{ 0 } ^{f _T } S _h ( f ) ( 2\pi f ) ^2 ~{\rm Re} C ( \theta _{ ab }, f )\,df
\label{eq:sigpb}
\end{equation} 
and
\begin{equation} 
\big\langle  j _{ {\rm GW}} ^{ ( a ) } j _{ {\rm GW}} ^{ ( b ) }  \big\rangle = \int _{ 0 } ^{f _T } S _h ( f ) ( 2\pi f ) ^4 ~{\rm Re}C ( \theta _{ ab }, f )\,df \,,
\label{eq:sigpd}
\end{equation} 
where we have used $ S _h ( -  f ) = S _h ( f )  $ to restrict the integrals to positive frequencies and use $ {\rm Re} C ( \theta _{ ab} , f ) $ to denote the real part of $ C ( \theta _{ a b } , f ) $. 

The function $ {\cal K} _{ ab } $ also appears in traditional pulsar searches, but its effects are typically neglected, setting this factor to unity. This is a good approximation for $ f d _a \gg 1 $ since $ {\cal K} _{ ab } $ is a highly oscillatory function, and contributions from the oscillating components are negligible. In this limit, the spatial correlations are frequency independent, and the final result is the well-known Hellings-Downs correlation matrix~\cite{1983ApJ...265L..39H},
\begin{align} 
\label{eq:hellings}
C ( \theta _{ ab }, f  )\rightarrow C ( \theta _{ ab} )  &  = x _{ ab }\log x _{ ab} - \frac{1}{6} x _{ ab } + \frac{1}{3} + \frac{1}{3} \delta _{ ab }\,,
\end{align} 
where $ x _{ ab } \equiv \frac{1}{2} ( 1 - \cos \theta _{ ab} ) $. In this limit, the acceleration and jerk correlations assume simplified forms,
\begin{align} 
\big\langle a _{ {\rm GW}} ^{ ( a ) } a _{ {\rm GW}} ^{ ( b ) } \big\rangle & \rightarrow  C(\theta_{ab})\int _0 ^{ f _T }S_h(f) (2\pi f ) ^2  df \hspace{0.25cm} {\rm and} \\ 
\big\langle j _{ {\rm GW}} ^{ ( a ) } j _{ {\rm GW}} ^{ ( b ) } \big\rangle & \rightarrow C(\theta_{ab})\int  _0 ^{ f _T }S_h(f) (2\pi f) ^4 df \,.
\end{align} 
 We exploit this simplification in Sec.~\ref{sec:peaked}.

In the ultralow-frequency regime, $ {\cal K} _{ a b} $ cannot always be set to unity. The exponential terms in $\mathcal{K}_{ab}$ cease oscillating for frequencies below either of the SSB-pulsar distances, resulting in a frequency-dependent correlator distinct from the Hellings-Downs average (see, e.g., Ref.\cite{Anholm:2008wy} for further discussion on this point). This effect is important for frequencies below the inverse distance to the nearest millisecond pulsar, which we denote $ f _d $. For the data sets we use, $ f _d  \simeq  100~{\rm pHz} $. 

\section{Methods}
\label{sec:methods}
We now use the results of the previous section to search for a stochastic gravitational-wave background. To this end, we carry out a log-likelihood ratio test using the acceleration and jerk correlators (Eqs.~\eqref{eq:sigpb} and \eqref{eq:sigpd}) to calculate the expected signal and compare it to the data. In this way, we are able to place strong constraints on the SGWB at sub-nHz frequencies, as described below.

\subsection{$ \dot{ P} _{ b} $ and $ \ddot{ P} $ Data Sets}
To search for signals in $ \dot{ P} _b $ and $ \ddot{ P} $, we use the same data sets as in Ref.~\cite{DeRocco:2022irl}. We briefly summarize the salient features of these data sets. Details of the pulsars used can be found in App.~\ref{app:datasets}.

The binary pulsar catalog used for the $ \dot{ P} _b$ data was initially compiled by Ref.~\cite{2021ApJ...907L..26C} to search for the Milky Way potential~\cite{10.1093/mnras/stv2395,10.1093/mnras/stw483,doi:10.1126/science.1132305,Fonseca_2014,Alam_2020,10.1093/mnras/sty2905,10.1111/j.1365-2966.2012.21253.x,Liu_2020,Cognard_2017} (see also Ref.~\cite{Phillips:2020xmf} for a similar analysis). Each of the 14 binary pulsars in the data set has estimates of the intrinsic and kinematic contribution to its observed $\dot{P}_b$ (first two terms of the right-hand side of Eq.~\eqref{eq:Pbd}). We estimate the contribution due to the Milky Way potential (third term of Eq.~\eqref{eq:Pbd}) using the \texttt{MWPotential2014} model implemented in the \texttt{galpy} Python package~\cite{Bovy_2015}. We assume a 20\% uncertainty on the value of $ a _{ {\rm MW}} $, in line with typical uncertainties of the galactic fit parameters in \texttt{MWPotential2014}. To estimate $ a _{ {\rm GW}} $ for each pulsar, we subtract away the intrinsic, kinematic, and galactic contributions to $ \dot{ P} _{ b, {\rm obs}} $. The dominant source of uncertainty changes from pulsar to pulsar, with the most sensitive pulsar (\verb|J1713+0747|) reaching an estimated acceleration of $ a _{ {\rm GW}} = ( 0.7 \pm 3.0 ) \times 10 ^{ - 20} \,{\rm sec} ^{-1}  $. 

The data set used for the $\ddot{P}$ analysis consists of 46 pulsars, whose observed $ \ddot{ P} $ were calculated in Ref.~\cite{Liu:2019iuh} using data from the EPTA~\cite{10.1093/mnras/stw483} and PPTA~\cite{10.1093/mnras/stv2395} collaborations. The authors of Ref.~\cite{Liu:2019iuh} included the effect of dispersion-measure (DM) variations and (high-frequency) red noise in their analysis, hence uncertainties deriving from them are already incorporated into the quoted uncertainty on the $\ddot{P}$ measurements. The DM spectrum was assumed to take the Kolmogorov turbulence form, with an amplitude as measured in Ref.~\cite{10.1093/mnras/sts486}. The red noise redshift spectrum, $ S _{ {\rm RN}} ( f ) $ was taken as a power-law, described by amplitude ($ A _{ {\rm RN}} $), high-frequency spectral index ($ \gamma $), and breaking frequency ($ f _ c $),
\begin{equation} 
S _{ {\rm RN}} ( f )  = \frac{A _{ {\rm RN}} ^2 }{3 f _{ \star}} \left( \frac{ f }{ f _\star } \right)   ^2  \left[ 1 + ( f / f _c ) ^2 \right] ^{ - \gamma /2 }\,.
\end{equation} 
where we define $ f_\star = 1~\text{yr}^{-1}$ as the reference frequency throughout. These parameters were marginalized over during the fitting procedure with prior estimates taken from EPTA and PPTA fits. 

The red noise was added by sampling the power spectrum at frequencies, $ n /T $, where $ n $ is an integer starting at $ 1 $~\cite{private}. This procedure does not account for the likely possibility that the red noise spectrum extends to frequencies below $ 1/ T $.\footnote{We thank the anonymous referee for alerting us to this subtlety.}  Such ultralow-frequency red noise is imprinted on the timing model, similar to the SGWB. To estimate the influence of such red noise, it is straightforward to use the formalism introduced in Sec.~\ref{sec:ultralow}. Assuming a redshift power spectrum, $ S _{ {\rm RN}}  ( f ) $, we find the red noise below $ 1/ T $ contributes a variance to $ \ddot{ P} $,
\begin{align} 
\sigma _{ {\rm RN}} ^2 \equiv \bigg\langle \bigg( \frac{ \Delta \ddot{ P} }{ P}\bigg) \bigg\rangle & =  \int _0 ^{ f _T } d f ~( 2\pi f ) ^4  ~S _{ {\rm RN}}( f ) \label{eq:RNPddot}\,.
\end{align} 
This serves as an additional source of Gaussian noise. We calculate the integral using the best-fit values from the EPTA~\cite{10.1093/mnras/stw483} and PPTA~\cite{10.1093/mnras/stv2395} datasets and add $ \sigma _{ {\rm RN}} $ in quadrature with the uncertainty on $ \ddot{ P } _{ {\rm obs}}/ P $.\footnote{Alternatively, an additional contribution, $ j _{ {\rm RN}} $, can be added to the right-hand side Eq.~\eqref{eq:Pdd} with $ \left\langle j _{ {\rm RN}} ^2 \right\rangle $ given by Eq.~\eqref{eq:RNPddot}. One can show that including a Gaussian prior on $ j _{ {\rm RN}} $ with a variance, $ \sigma _{ {\rm RN}} ^2 $, is equivalent to adding $ \sigma _{ {\rm RN}} $ in quadrature with the observed standard deviation.} For pulsars fit by both collaborations, we use the estimate in Eq.~\eqref{eq:RNPddot}, which yields a larger value. Since $ \sigma _{ {\rm RN}} $ depends on $ f _T $, it contains a significant degree of uncertainty. To account for this, we also present results which inflate $ \sigma _{ {\rm RN}} $ by an order of magnitude. We note that, in principle, a similar expression applies in the $ \dot{ P} _b $, however, in that case, the analogous estimate to that in Eq.~\eqref{eq:RNPddot} is negligible We display the values of $\sigma_{\text{RN}}$ for each pulsar in Tab.~\ref{tab:Pddot}.

 Three pulsars in the set (\verb|J1024-0719|, \verb|B1821-24A|, and \verb|B1937+21|) are poorly suited for a search for gravitational-wave signals due to a wide binary companion, location within a dense cluster, or significant ultralow-frequency red noise. We omit these three from our analysis. The most sensitive pulsar in this search is \verb|J0613-0200| and provides an estimate of $ j _{ {\rm GW}} =  ( 0.6 \pm 0.6 ) \times 10 ^{ - 30 }\, {\rm sec} ^{ - 2} $. 

\subsection{Statistical Analysis}
\label{sec:stats}
To carry out the analysis, we construct a likelihood individually for each data set, assuming uncertainties between the measurements are uncorrelated. We denote the data by $ \left\{ s _{ {\rm GW}} \right\} = \left\{ a _{ {\rm GW}} \right\} $ or $ \left\{ j _{ {\rm GW}} \right\} $, the signal model for all the pulsars with a bar ($ \left\{ \bar{ s} _{ {\rm GW}} \right\}  = \big\{ \bar{ a}_{\text{GW}} ^{ ( a ) } \big\} $ or $\big\{ \bar{ j}_{\text{GW}} ^{  ( a )} \big\} $), and the uncertainty in the measurement as $ \sigma _a $.~\footnote{The uncertainty on the measurement includes the red noise contribution, $ \sigma _{ {\rm RN}} $, mentioned above.} The likelihood is:~\footnote{In our analysis, the influence of gravitational waves is in the offsets, $ \bar{ s} _{ {\rm GW}} ^{( a ) } $. One could alternatively introduce a Gaussian prior on the model prediction with a covariance determined by the two-point correlator derived in the main text and integrate over $ \bar{ s} _{ {\rm GW}} ^{(a)}  $. This would result in a Gaussian likelihood for $ s _{ {\rm GW}} ^{ ( a ) } $ with mean 0, but a non-diagonal covariance matrix.}
\begin{align}
{\cal L} (\bar{ s } _{ {\rm GW}}   | \left\{ s _{ {\rm GW}}   \right\} ) & = \prod _a \frac{1}{\sqrt{ 2\pi \sigma _a^2  }} \exp\left[-\frac{1}{ 2 \sigma _a^2}( s _{ {\rm GW}} ^{ (  a ) } - \bar{ s} _{ {\rm GW}} ^{ ( a ) } ) ^2  \right] \,. \label{eq:likelihood}
\end{align}

The model prediction, $ \left\{ \bar{ s} _{ {\rm GW}} \right\} $, is a random vector of length $ 14$ (46) for the $\dot{P}_b$ ($\ddot{P}$) analysis respectively. For each analysis, we average the likelihood over 10,000 realizations of the vector, each drawn from a multivariate Gaussian distribution with a correlator given by Eq.~\eqref{eq:sigpb} or Eq.~\eqref{eq:sigpd}. The averaging procedure yields a marginalized likelihood that, due to the presence of $S_h(f)$ in the correlators from which the mock data was sampled, is implicitly a function of the GW spectrum. We denote this marginalized likelihood $\mathcal{L}_\text{M}$. We compute the log-likelihood ratio using $ {\cal L} _M $ between a signal model specified by the dimensionless normalization of $S_h(f)$ (denoted $S_0$) and the null hypothesis of zero GW signal ($S_0 = 0$):
\begin{equation} 
\hat{q}(S _0 )   = -2 \log \left( \frac{ \mathcal{L} _{ {\rm M}} ( S_0 | \left\{ s _{ {\rm GW}}  \right\}) }{ \mathcal{L} _{ {\rm M}} (S _0  = 0 | \left\{ s _{ {\rm GW}} \right\})} \right) \,.
\end{equation} 
We assume this statistic obeys Wilks' theorem for a parameter bounded on one side by zero~\cite{Algeri_2020} to set the 95\% confidence limit on $S_0$ at $\hat{q}(S_0)  = 2.71$. This assumption has been confirmed in the case of a continuous wave source in Ref.~\cite{DeRocco:2022irl}.

\section{Applications and Results}
\label{sec:applications}

As with any gravitational-wave analysis, a bound on the stochastic background can only be set once one specifies a power spectrum, $ S _h ( f ) $. In this section, we consider three types of gravitational-wave distributions and set limits on each. The first is a monochromatic distribution, which is particularly useful for elucidating spectral dependence of the sensitivity. The second is a bounded distribution with support only at frequencies exceeding $ f _d $. Finally, we consider power-law distributions, such as that expected from supermassive black hole binaries, and compare our limits on these distributions to the tentative signal currently observed at higher frequencies. 

\subsection{Monochromatic analysis}
\label{sec:mono}
In this subsection, we place bounds on the relic energy density of a monochromatic spectrum at a given frequency $f$,
\begin{equation} 
\label{eq:mono}
S _h  ( f ') = S _0\,\delta ( f '- f   ) \,.
\end{equation} 
The monochromatic spectrum illuminates the frequency-dependence of our methodology, hence is useful for comparison to the expected spectra of various signals. The total energy density corresponding to Eq.~\eqref{eq:mono} is given by the integral,
\begin{align}
\label{eq:omega}
\Omega_\text{GW}= \int _0 ^{\infty} \frac{ d f' }{ f' }  \frac{ 4\pi ^2 f ^{\prime 3} }{ 3 H _0 ^2 } S _h ( f' ) & = \frac{4\pi ^2 f ^2   S _0}{3 H _0 ^2}\,.
\end{align}
For each $ f  $, we set a bound on $  S _0  $, which we translate into a bound on the total energy density that could be contributed by a monochromatic distribution of GWs at $f$ using Eq.~\eqref{eq:omega}. We use the bound on the total energy density, $ \Omega _{ {\rm GW}} $, as a proxy for the bound on the differential energy density per unit log frequency, $ \Omega _{ {\rm GW}} ( f  ) $.

At frequencies above 1~nHz, pulsar timing collaborations perform a similar estimation via ``free-spectrum analysis,''  where the spectral power is taken as a sum over $N$ monochromatic contributions with frequencies $f_n = n/T; n=1,2,...\,N$~\cite{Lentati:2015qwp,Shannon:2015ect,NANOGRAV:2018hou}. Here $T$ is the longest observation time for a pulsar in the dataset. For each $ f _n $, they study the bound on the corresponding amplitude, allowing all other amplitudes to vary simultaneously and marginalize over their posteriors. A bound on the amplitude is used as a proxy for the bound on the characteristic strain, which can be translated into a bound on $ \Omega _{ {\rm GW}} ( f ) $. Note that while both our monochromatic methodology and the free-spectrum analysis used by PTA collaborations are effective proxies for the spectral energy density, they are not formally equivalent to one another. We have chosen to plot the PTA curves on our Fig.~\ref{fig:sensitivity} to provide a qualitative comparison to our results.

As discussed in Sec.~\ref{sec:ultralow}, for our methodology to apply, the signal must resemble a slow secular drift. We ensure this by restricting our analysis to frequencies below $f_T $. For the $ \dot{ P } _b $ ($ \ddot{ P} $) data set, $ T _{ {\rm max}} = 22 $ ($ 17.7 $) yr, corresponding to $ f _T = 360~{\rm pHz}$ ($ 466~ {\rm pHz} $). At frequencies above $ f _T $, oscillatory features of the GW signal are important, and instantaneous derivatives of $ v _{ {\rm GW}} $ no longer determine the impact to the timing model.

The limits on $\Omega_\text{GW} ( f ) $ set by the $ \ddot{ P} $ and $ \dot{ P} _b $ analyses are shown in Fig.~\ref{fig:sensitivity} (red) alongside free-spectrum analyses performed by the pulsar timing collaborations~\cite{Lentati:2015qwp,Shannon:2015ect,NANOGRAV:2018hou} (blue).\footnote{These limits are from 2015-2018 and hence predate the recent evidence of a signal. The more recent analyses are no longer setting limits and hence are not directly relevant for limit setting in Fig.~\ref{fig:sensitivity}. We will compare our results with the signal observation in Sec.~\ref{sec:powlaw}.} The theoretical expectation from SMBH binaries is a broken power-law spectrum with a known index at high frequencies and a spectral index at low frequencies sensitive to an undetermined energy loss mechanism around separations of 1~pc.\footnote{See Ref.~\cite{Dror:2021wrl} for a discussion of possible modifications to this spectrum due to physics beyond the Standard Model.} Following Ref.~\cite{Sampson:2015ada}, we parameterize the spectrum as,
\begin{equation} 
\label{eq:smbh}
S _h ( f ) = \frac{ A _{ \star} ^2 } { 2 f _{ \star}} \frac{ ( f / f _{\star} ) ^{ 2 - \gamma }} { 1 + ( f _b / f ) ^\kappa  }\,,
\end{equation} 
where $ f _b $ is an unknown ``bending'' frequency expected to be well below 1~nHz, $ \kappa = 10/3 $ if stellar scattering dominates the energy losses at large binary separations,\footnote{The dominant energy loss mechanism during this stage has been a topic of active discussion in the literature and is often referred to as the ``final parsec problem''~\cite{Milosavljevic:2002ht} (see, e.g., Refs.~\cite{2011ApJ...732L..26P,2011ApJ...732...89K,2016MNRAS.463..858P,2015ApJ...810...49V,10.1093/mnras/stw2528,10.1093/mnras/sty3458,Gualandris_2022} for discussions on possible resolutions and Refs.~\cite{Sampson_2015,Taylor_2017} for discussion of how it may be probed by PTA analyses at frequencies above 1 nHz).} and $ \gamma = 13/3 $ if gravitational-wave emission dominates the energy loss at small separations. Recent results from NANOGrav find a best-fit of the amplitude of, $ A _{ \star} = 2.4 ^{ + 0.7} _{ - 0.6} \times 10 ^{ - 15} $ for $ \gamma = 13/ 3 $~\cite{NANOGrav:2023gor}. We show the spectrum for $ \kappa = 10/3 $, $ \gamma = 13/3 $, $ A _{ \star} $ given by the NANOGrav best-fit value, and different bending frequencies in Fig.~\ref{fig:sensitivity} (orange). We also show the best fit results of NANOGrav's free-spectrum analysis in gray~\cite{NANOGrav:2023gor}. 

The $ \ddot{ P} $ sensitivity scales with the square of the frequency for frequencies well above the inverse SSB-pulsar distances while the $ \dot{ P } _b $ sensitivity is approximately constant. This can be understood from the scaling of the apparent velocity in the monochromatic regime. From Eq.~\eqref{eq:sigpb}, $ \langle a_\text{GW} ^2 \rangle  \propto S _0  f _0 ^2 $ and $ \langle j _{ {\rm GW}} ^2 \rangle \propto S _0 f _0 ^4 $. Since $ S _0 $ is proportional to energy density over frequency squared (see Eq.~\eqref{eq:omega}), holding $a_\text{GW}$ or $j_\text{GW}$ fixed yields $\Omega _{ {\rm GW}} \propto f^0$ ($ f ^{ - 2} $) for an $a_\text{GW}$ ($j_\text{GW}$) search.

Below frequencies of $ {\cal O} ( 10~{\rm pHz} ) $, the wavelength is longer than the typical SSB-pulsar distance of 1~kpc. As a result, the SSB and pulsar experience correlated motion that partially cancels the signal. In the $ \ddot{ P} $ analysis, this results in the bend at approximately 10~pHz. In the $ \ddot{ P} _b $ analysis, the sensitivity is largely driven by two pulsars, both with distances of around 1~kpc.\footnote{A few pulsars in the analysis are substantially closer with SSB-pulsar distance leading to the features observable at $ 100~{\rm pHz} $. Since these do not drive the sensitivity, passing through this threshold does not substantially change our limits.} This results in a noticeable bump in the sensitivity at the corresponding frequency due to the non-trivial functional form of $\mathcal{K} _{ ab } ( f , \hat{{\bf n}} )  $ (Eq.~\eqref{eq:K}) and a subsequent bend in the sensitivity at even lower frequencies.

While the $ \ddot{ P} $ analysis is significantly more sensitive than the $ \dot{ P} _b $ analysis above a few pHz, the two serve as complementary probes of a signal. If a signal is observed with both parameters, a joint search will provide significant insight into the spectral shape of the ultralow-frequency SGWB.

\subsection{Distributions with $ f \gtrsim  f _{ d } $}
\label{sec:peaked}
If a GW spectrum has support dominantly at frequencies greater than the inverse SSB-pulsar distance ($ f \gg f _{ d } \simeq  100\,{\rm pHz} $), the $ a _{ {\rm GW}} $ and $ j _{ {\rm GW}} $ correlators (Eqs.~\eqref{eq:sigpb} and \eqref{eq:sigpd}) are separable into frequency-dependent and angular-dependent components. In this limit, the frequency integral sets the overall amplitude of the correlator and scales with the normalization. Consequently, the entire integral can be constrained by following the procedure in Sec.~\ref{sec:stats}.

Performing this analysis, we find
\begin{align} 
 \int _0 ^{f _T   }d f ( 2\pi f ) ^2 S _h ( f )   & <  1.0 \times 10  ^{ -38}\, {\rm sec} ^{ - 2 }\quad {\rm and}\label{eq:limPbd}\\ 
\int  _0 ^{ f _T   }d f ( 2\pi f ) ^4 S _h ( f )   & <  1.6 \times 10  ^{ -61}\, {\rm sec} ^{ - 4 } \label{eq:limPdd}\,.
\end{align} 
The powers of frequency within the integrals in Eq.~\eqref{eq:limPbd} and \eqref{eq:limPdd} show a general feature that the sensitivity of the $ \ddot{ P} $ and $ \dot{ P} _b $ analysis will be most sensitive to the highest frequencies unless the spectrum falls rapidly with increasing frequency. While these results are only applicable to distributions with support in a narrow range of frequencies, $ 100~{\rm pHz} \lesssim f \lesssim 450~{\rm pHz} $ (the upper bounds coming from $ f _T $, see previous subsection), the expressions apply for \textit{any} such distribution, $ S _h ( f ) $. 

\subsection{Power law distributions}
\label{sec:powlaw}
The expected stochastic signal from supermassive black hole mergers or cosmological sources is often well approximated by a power law in the most sensitive experimental frequency range, 
\begin{equation} 
\label{eq:powlaw}
S _h ( f ) = \frac{ A _\star ^2 }{ 2 f _\star  } \left( \frac{ f }{ f _{\star} } \right) ^{2 - \gamma}\,.
\end{equation} 
 For the SMBH spectrum discussed in Sec.~\ref{sec:mono}, this is a reasonable approximation when contributions from $ f \lesssim f _b $ are negligible. For the $ \dot{ P} _b $ ($ \ddot{ P} $) analysis, this tends to be the case when $ \gamma <5 $ ($ <7$), both of which are satisfied by the theoretical expectation of $\gamma = 13/3$. By substituting $S_h(f)$ from Eq.~\eqref{eq:powlaw} into the correlators in Eqs.~\eqref{eq:sigpb} and \eqref{eq:sigpd}, we can carry out the analysis described in Sec.~\ref{sec:stats} to produce limits on $A_\star$ and $\gamma$.

\begin{figure} 
  \includegraphics[width=8cm]{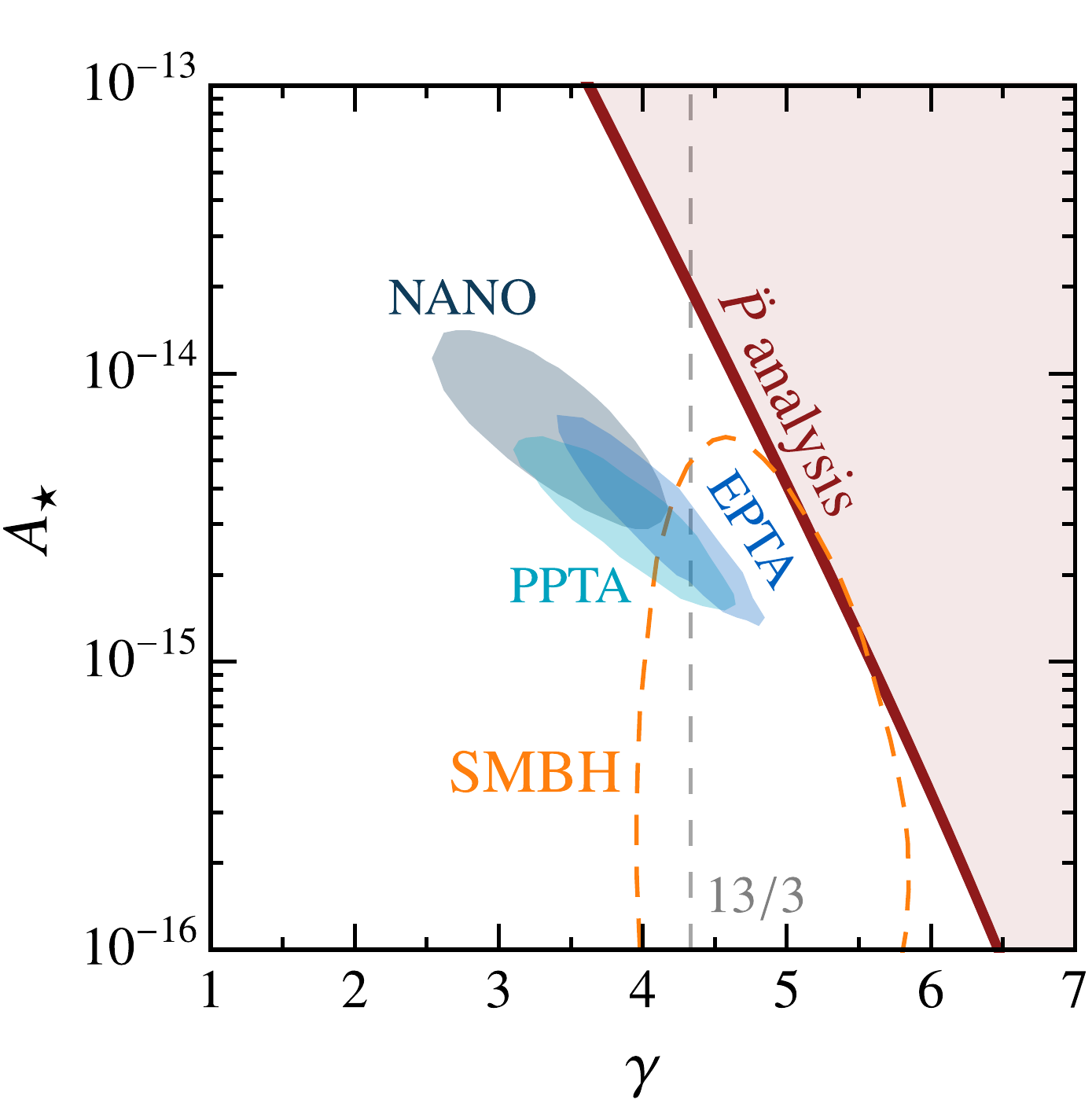}
\caption{Comparison of the tentative signal found at frequencies above 1~nHz as fit by NANOGrav~\cite{NANOGrav:2023gor}, EPTA~\cite{EPTA:2023fyk}, and PPTA~\cite{Reardon:2023gzh} to the limits from the $ \ddot{ P} $ analysis ({\bf \color{Ourcolor}red}) performed in Sec.~\ref{sec:powlaw}. The width of the line in the $ \ddot{ P}  $ limit corresponds to taking the ultralow-frequency red noise, $ \sigma _{ {\rm RN} } $, from its estimated value to ten times the estimated value (see text for details). For comparison, we also show the range of $ \left\{ A , \gamma \right\} $ values found from a population study of supermassive black holes~\cite{Agazie_2023}.}
\label{fig:Agamma}
\end{figure}

The resulting limits from a $ \ddot{ P} $ analysis are shown in Fig.~\ref{fig:Agamma} as a red curve. Limits from $\dot{P}_b$ are inferior to those from $ \ddot{ P} $ except for very large values of $ \gamma $. For $ \gamma < 7 $, the sensitivity of the $ \ddot{ P } $ analysis is driven by frequencies near $ f _T $. In this case, by inserting the spectrum in Eq.~\eqref{eq:powlaw} into Eq.~\eqref{eq:limPdd} one can see the approximately linear scaling of $\log A_\star$ with $\gamma$. The blue regions denote the $ 2 \sigma $ best-fit confidence intervals of the signal from NANOGrav~\cite{NANOGrav:2023gor}, EPTA~\cite{EPTA:2023fyk}, and PPTA~\cite{Reardon:2023gzh}. For comparison, we also show the expected range of predictions of $ \left\{ A, \gamma \right\} $ from simulations of supermassive black hole mergers performed with \texttt{Holodeck} simulations~\cite{Agazie_2023}. We conclude that doing a global analysis using both ultralow and higher frequencies would disfavor large values of $ \gamma $.

For supermassive black hole binaries, the stochastic gravitational-wave signal is expected to be within reach of existing PTA analyses. Importantly, it may be the source of the common-process signal that the PTA collaborations are currently detecting at an amplitude $A_\star \simeq  2.8 ^{ + 1.2} _{ - 0.8} \times 10 ^{ - 15}$~\cite{Antoniadis:2022pcn}. Our results place constraints near this best-fit value, limiting $A_\star < 1.8 \times 10^{-14}$ at $ \gamma = 13/3 $.

\section{Discussion}
\label{sec:discussion}
A stochastic background of gravitational waves below 1~nHz is well-motivated by the expected signal from supermassive black hole mergers and potential cosmic sources. In this paper, we have shown that the secular drift of parameters in pulsar timing models can detect such an SGWB. We do not find significant evidence of a signal in existing data, and hence we place strong constraints on the SGWB spectrum. Our key results are as follows:
\begin{enumerate}
\item We limit the relic energy density for a monochromatic spectrum at 450~pHz to $\Omega_\text{GW} ( f )  < 3.8 \times 10 ^{ - 9} $, with sensitivity which scales as the frequency squared until approximately 10~pHz.
\item For spectra with dominant support between $ 100~{\rm pHz} \lesssim f \lesssim 450~{\rm pHz} $, we put a generic constraint on integrals over the power spectral density, $ S _h ( f ) $. 
\item We limit the amplitude of power-law spectra as a function of a spectral index. For $ \gamma = 13 /3 $ (expected by SMBH mergers), we find $A_\star \lesssim 1.8 \times 10^{-14}$ for a reference frequency of $f_* = \text{year}^{-1}$.
\end{enumerate}
While our study already significantly influences the prospects for searching for an SGWB, the data used in the $ \ddot{ P} $ analysis are not from the most recent EPTA and PPTA data sets, nor do they include any NANOGrav observations. An updated analysis with the new data releases might detect the SGWB at frequencies below 1~nHz. This regime is an important complementary GW probe to conventional PTA analyses. Furthermore, it would provide key insights into the spectral shape of the SGWB that are only possible in the ultralow-frequency regime. In the case of SMBH mergers, the ultralow-frequency signal is sensitive to the bending frequency $f_b$ of the spectrum (Eq.~\eqref{eq:smbh}). This would provide insight into the undetermined physics driving the merger at these scales.

Whether or not a sub-nanohertz detection is made by studying the pulsar timing model parameters using the most recent PTA data, we know that the SGWB \textit{is} out there and is very likely observable in the near future. As such, the methodology described in this paper provides a guaranteed detection in the years to come, opening an entirely new frequency range of the gravitational-wave spectrum to explore.

\vspace{0.2cm}
\noindent {\it Acknowledgements.}
The authors thank Steve Taylor and the GW group at Vanderbilt University for insight during the writing of this paper, Xiao-Jin Lui for communications about the procedure used in Ref.~\cite{Liu:2019iuh}, and the anonymous referee for suggestions on incorporating the ultralow-frequency red noise as well as constructive comments on the manuscript. WD would like to thank Daniel Egana-Ugrinovic for lively discussion on the neutron magnetosphere. JD would like to thank the Indian Pulsar Timing Array collaboration for enlightening discussions. The research of JD is supported in part by NSF CAREER grant PHY-1915852 and in part by the U.S. Department of Energy grant number DE-SC0023093. WD is supported in part by Department of Energy grant number DE-SC0010107. Part of this work was performed at the Aspen Center for Physics, which is supported by National Science Foundation grant PHY-1607611.

\bibliographystyle{JHEP}
\bibliography{refs}

\onecolumngrid

\appendix
\count\footins = 1000

\section{Non-Stationary Residuals}
\label{app:residuals}
In the main body of the text, we assumed the timing model is sufficiently extensive to capture all secular variation in the time of arrival data. If this is not the case, e.g., the model does not fit $\ddot{P}$ in the presence of ultralow-frequency signals, then a secular trend persists in the residuals. In this section, we compute the correlator of the residuals in this case.

The residuals, which we denote as $ R _n   ( \lambda    ) $, are given as the difference between the arrival time of the $ n $th pulse and the expected time of arrival given the timing model, $ \bar{t} _n ( \lambda ) $, where $\lambda$ denotes the model parameters. There are two contributions to this difference. The first is from the mismatch of the timing model with the true parameters. The second is from a GW signal and is equal to an integral over the induced redshift, $ z _a  ( t ) $ (denoted by $ v _{ {\rm GW}} ^{ ( a ) }$ in the main text). Normally, it is assumed that the pulsar parameters have already been measured with sufficient accuracy that we can expand about small deviations from these known values. In this limit,
\begin{equation} 
\label{eq:res}
R _n ( \lambda  )  \simeq M _{ n \alpha } ( \lambda _\alpha - \hat{\lambda} _\alpha )    +   \int _0 ^{t _n } d t' z ( t ' ) \,,
\end{equation}  
where $ \hat{\lambda} _\alpha  $ denotes the true value of $ \alpha $th parameter, and $ M _{ n \alpha } $ is commonly known as the design matrix. Now suppose that the timing model is chosen to incorporate $ P $ and $ \dot{ P} $, but does not fit $ \ddot{ P} $. In this case, the signal is embedded in the integral over redshift. We now compute its corresponding covariance matrix. 

Using Eq.~\eqref{eq:vEP} and a Fourier decomposition of the gravitational field, $\tilde{h}$, the redshift can be written as 
\begin{equation} 
z _a ( t ) = \sum _{ A = + , \times } \int  _{ - \infty } ^{\infty} d f \int d ^2 \hat{{\bf n}} \,\tilde{h} _A ( f , \hat{{\bf n}} ) F _a ^A e ^{ - 2\pi i f t }\left[ 1 - e ^{ 2\pi i f d _a ( 1 + \hat{{\bf n}} \cdot \hat{{\bf n}} _a ) }\right] \,.
\end{equation} 
From here we can compute the ensemble average of a product of redshifts at two different times,
\begin{equation} 
\left\langle z _a ( t ) z _b  ( t ' ) \right\rangle = \frac{1}{2} \int _{ - \infty }^{\infty} d f S _h  ( f ) C ( \theta _{ ab } , f ) e ^{ - 2\pi i f ( t - t ' ) } \,,
\end{equation} 
where $ C ( \theta _{ a b }, f  ) $ is the overlap function introduced previously which reduces down the frequency-independent form at frequencies much greater than $ d _a ^{-1} $ and $ d _b ^{-1} $. As usual, this correlator should be interpreted as an ensemble average.

The correlator of the residuals is then given by
\begin{align} 
\left\langle R _a ( t ) R _b  ( t ' ) \right\rangle & = \frac{1}{2} \int _{ - \infty } ^{\infty} d  f \frac{S _h  ( f  ) }{ ( 2\pi f )  ^2 } \left( 1 - e ^{ 2\pi i f  t } \right) \big( 1 - e ^{ - 2\pi i f   t ' } \big) C ( \theta _{ ab } , f ) \,.\label{eq:rescorr}
\end{align} 
This equation is the general form of the correlator that must be used if one wants to study ultralow-frequency GWs. Clearly, it is not stationary; the residual two-point function depends on $ t $ and $ t' $ explicitly. To search for gravitational waves both above and below $ 1/T $, one must include the full expression in Eq.~\eqref{eq:rescorr}. \footnote{Note that in an analysis which targets frequencies such that $2\pi f  t \gg 1 $, $ C ( \theta _{ a b } , f ) $ becomes real and the correlator takes a simplified form. In this limit,
$$
\left\langle R _a  ( t ) R _b  ( t ' ) \right\rangle  \simeq   C ( \theta _{ ab } )  \int _0 ^{\infty} d f  \frac{ S _h  ( f  ) }{ ( 2\pi f )  ^2 }\left( 1  + \cos ( 2\pi f  ( t - t ' ) ) \right) \,,
$$
where we have dropped contributions from terms in the integrand proportional to $ \cos ( 2\pi f t ) $ or $ \cos ( 2\pi f t ' ) $ as they are highly oscillatory and subdominant for high-frequencies. The time-independent piece represents overall shifts in the residuals, and is assumed to be difficult to measure. Dropping this piece gives the form typically found in the PTA literature (see, e.g., Ref.~\cite{Hazboun:2019vhv}).}

\section{Analysis Data Sets}
\label{app:datasets}
In this work we carried out two analyses with the pulsar timing model parameters, one using $ \dot{ P} _b $ and another using $ \ddot{ P} $. In this appendix, we provide tables of the parameter values used in each analysis. An extensive discussion of the selection criteria and validation of these data sets can be found in App. S-IB of Ref.~\cite{DeRocco:2022irl}.

The $ \dot{ P} _b $ analysis was carried out using the data for the 14 pulsars with binary companions shown in Tab.~\ref{tab:Pbdot}. The data was initially compiled by Ref.~\cite{2021ApJ...907L..26C}. We chose this set since each pulsar's intrinsic and kinematic contributions were already estimated via independent measurements.~\footnote{This reference did not estimate the intrinsic contribution to pulsars J0613-0200, J1614-2230, and J1713+0747; we used the quadrupole radiation formula to estimate the intrinsic value.} The contributions from acceleration in the Milky Way potential were estimated via \texttt{MWPotential2014}, as described in the main body of the text. 

In Tab.~\ref{tab:Pddot}, we show the data for the 46 pulsars used in the $ \ddot{ P } $ analysis. The $ \ddot{ P} $ values for each pulsar were calculated by Ref.~\cite{Liu:2019iuh}. In addition to the pulsars we list in this table, Ref.~\cite{Liu:2019iuh} studied $\ddot{P}$ for three other pulsars: J1024-0719, B1821-24A, and B1937+21. J1024-0719 is believed to be in a wide binary orbit with a period between 2000-20000 years that induces a large anomalous $ \ddot{ P} $~\cite{Kaplan:2016ymq}, while B1821-24A is situated in a dense cluster in which gravitational effects due to nearby stars are non-negligible~\cite{Bilous_2015}. The residuals of B1937+21 are subject to significant ultralow-frequency red noise relative to its timing precision~\cite{1994ApJ...428..713K}.  We, therefore, exclude these pulsars from our data set. 

\begin{table}
\begin{center}
\caption{Pulsars used for $\dot{P}_b$ analysis. $( l, b ) $ is the galactic longitude and latitude, $d _a $ is the distance between the Earth and the pulsar ($ a $), $T$ is the observation time, $\dot{P}_{b,\text{obs}}/P_b$ is the observed value of the line-of-sight acceleration, $ \dot{P}_{b,\text{int}}/ P_b$ is the intrinsic relative change in the binary period induced by gravitational emission, $v _{ \perp } ^2  / d _L $ is the estimated contribution from Earth-pulsar proper motion, $a_{\text{MW}}$ is the estimated contribution from Galactic accelerations, $\Delta a$ is the leftover contribution to the orbital pulsar derivative when the prior three are subtracted from $a_{\text{obs}}$, and Ref. is the reference from which we extracted the parameters. All contributions to $ \dot{P}_{b}/P_b $ listed below are in units of $10^{-18}$ s$^{-1}$. Pulsars for which the intrinsic contribution has been estimated using the quadrupole approximation to binary radiation are demarcated with a $\dagger$ (*) with inputs taken from Ref.~\cite{Reardon_2021} (\cite{Fonseca_2016}).}
\label{tab:Pbdot}
\begin{tabular}{lllllllllll}
\toprule
Pulsar &    $l$ (deg) &     $b$ (deg) &          $d _a $ (kpc) &     $T$ (yr) &       $\dot{P}_{b,\text{obs}}/P_b$ &           $  \dot{P}_{b,\text{int}}/ P_b$ &      $ v _{ \perp  } ^2  / d _L  $ &      $a_\text{MW}$ &     $\Delta a$ & Ref. \\
\midrule
J0437-4715   &  253.39 & -41.96 &  0.1570(22) &   4.76 &    7.533(12) &   -0.00552(10) &   7.59(11) &   -0.055(11) &   0.0(1) &             \cite{10.1093/mnras/stv2395} \\
J0613-0200   &  210.41 &  -4.10 &     0.80(8) &  16.10 &     0.46(11) &      -$0.02(5)^\dagger$ &  0.215(22) &     0.046(9) &    0.2(1) &              \cite{10.1093/mnras/stw483} \\
J0737-3039AB &  245.24 &  -4.50 &    1.15(18) &   2.67 &  -142.0(1.9) &   -141.565(15) &  0.053(16) &   -0.056(11) &   -0.5(19) &       \cite{doi:10.1126/science.1132305} \\
J0751+1807   &  202.73 &  21.09 &    1.22(25) &  17.60 &    -1.54(11) &      -1.91(17) &   0.56(12) &    0.048(10) &   -0.24(23) &              \cite{10.1093/mnras/stw483} \\
J1012+5307   &  160.35 &  50.86 &    1.41(34) &  16.80 &      1.17(8) &    -0.1955(33) &     2.3(5) &   -0.070(14) &     -0.8(5) &              \cite{10.1093/mnras/stw483} \\
J1022+1001   &  231.79 &  51.50 &   0.719(21) &   5.89 &     0.82(34) &    -0.0021(19) &  0.512(15) &   -0.130(26) &    0.4(3) &             \cite{10.1093/mnras/stv2395} \\
J1537+1155   &   19.85 &  48.34 &    1.16(24) &  22.00 &    -3.766(8) &     -5.3060(8) &     1.8(4) &     -0.19(4) &     -0.09(38) &                      \cite{Fonseca_2014} \\
J1603-7202   &  316.63 & -14.50 &      0.9(7) &   6.00 &     0.57(28) &         0.0(0) &   0.13(10) &    -0.039(8) &    0.5(3) &             \cite{10.1093/mnras/stv2395} \\
J1614-2230   &  352.64 &   2.19 &     0.65(4) &   8.80 &     2.10(17) &   -0.000558(5)* &   1.66(12) &    0.079(16) &    0.4(2) &                         \cite{Alam_2020} \\
J1713+0747   &   28.75 &  25.22 &     1.15(5) &  21.00 &    0.058(26) &  -1.03(6)e-06$^\dagger$ &   0.111(5) &   -0.060(12) &  0.007(28) &             \cite{10.1093/mnras/sty2905} \\
J1738+0333   &   27.72 &  17.74 &    1.47(11) &  10.00 &    -0.56(10) &       -0.91(6) &  0.270(20) &  -0.0049(10) &    0.09(12) &  \cite{10.1111/j.1365-2966.2012.21253.x} \\
J1909-3744   &  359.73 & -19.60 &   1.161(18) &  15.00 &   3.8645(10) &   -0.02111(23) &    3.88(6) &     0.034(7) &     -0.02(6) &                          \cite{Liu_2020} \\
J2129-5721   &  338.01 & -43.57 &    0.53(25) &   5.87 &       1.4(6) &         0.0(0) &    0.19(9) &   -0.121(24) &      1.3(6) &             \cite{10.1093/mnras/stv2395} \\
J2222-0137   &   62.02 & -46.08 &  0.2672(11) &   4.00 &       0.9(4) &    -0.0365(19) &   1.324(5) &   -0.098(20) &     -0.2(4) &                      \cite{Cognard_2017} \\
\bottomrule

\end{tabular}
\end{center}
\end{table}

\begin{table}
\begin{center}
\caption{Pulsars used for $\ddot{P}$ analysis. $l$ is galactic longitude, $b$ is galactic latitude, $d$ is the distance between the Earth and the pulsar, $T$ is the observation time, $\ddot{P}_\text{obs}/P $ is the observed line-of-sight jerk, $\sigma_\text{RN}$ is the additional uncertainty due to ultralow-frequency red noise (see Sec.~\ref{sec:methods}), and Ref. is the reference with we extracted the pulsar parameters.}
\label{tab:Pddot}
\begin{tabular}{llllllll}
\toprule
Pulsar &       $ l$ (deg) &       $b$ (deg) &      $d$ (kpc) &    $T$ (yr) &        $\ddot{P}_\text{obs}/P$ ($10^{-30}$ s$^{-2}$) &  $\sigma_\text{RN}$  ($10^{-30}$ s$^{-2}$) & Ref. \\
\midrule
J0030+0451 &  113.141 & -57.611 &  0.324 &  15.1 &          -4(4) &    0.54       &  \cite{10.1093/mnras/stw483} \\
J0034-0534 &  111.492 & -68.069 &  1.348 &  13.5 &          0(20) &           0 & \cite{10.1093/mnras/stw483} \\
J0218+4232 &  139.508 & -17.527 &  3.150 &  17.6 &          -2(5) &            0.14& \cite{10.1093/mnras/stw483} \\
J0437-4715 &  253.394 & -41.963 &  0.157 &  14.9 &      -1(1) &            0  & \cite{10.1093/mnras/stv2395} \\
J0610-2100 &  227.747 & -18.184 &  3.260 &   6.9 &       0(50) &           0 & \cite{10.1093/mnras/stw483} \\
J0613-0200 &  210.413 &  -9.305 &  0.780 &  16.1 &         0.6(6) &         0.13   & \cite{10.1093/mnras/stw483} \\
J0621+1002 &  200.570 &  -2.013 &  0.425 &  11.8 &        -70(30) &         1.28   & \cite{10.1093/mnras/stw483} \\
J0711-6830 &  279.531 & -23.280 &  0.106 &  17.1 &       1(1) &             0 & \cite{10.1093/mnras/stv2395} \\
J0751+1807 &  202.730 &  21.086 &  1.110 &  17.6 &       0(2) &            0 & \cite{10.1093/mnras/stw483} \\
J0900-3144 &  256.162 &   9.486 &  0.890 &   6.9 &        -10(20) &          0  & \cite{10.1093/mnras/stw483} \\
J1012+5307 &  160.347 &  50.858 &  0.700 &  16.8 &         0.4(7) &          0.01  & \cite{10.1093/mnras/stw483} \\
J1022+1001 &  231.795 &  51.101 &  0.645 &  17.5 &      -2(1) &           0.01  & \cite{10.1093/mnras/stw483} \\
J1045-4509 &  280.851 &  12.254 &  0.340 &  17.0 &          -2(7) &          0.05   & \cite{10.1093/mnras/stv2395} \\
J1455-3330 &  330.722 &  22.562 &  0.684 &   9.2 &          6(20) &            0.17 & \cite{10.1093/mnras/stw483} \\
J1600-3053 &  344.090 &  16.451 &  1.887 &   9.1 &           4(5) &             0.06 & \cite{10.1093/mnras/stv2395} \\
J1603-7202 &  316.630 & -14.496 &  0.530 &  15.3 &           1(4) &             0.02 & \cite{10.1093/mnras/stv2395} \\
J1640+2224 &   41.051 &  38.271 &  1.515 &  17.3 &        -0.9(9) &            0 & \cite{10.1093/mnras/stw483} \\
J1643-1224 &    5.669 &  21.218 &  0.740 &  17.3 &      -2(2) &           0.01  & \cite{10.1093/mnras/stw483} \\
J1713+0747 &   28.751 &  25.223 &  1.311 &  17.7 &        -0.5(5) &      0.38    &   \cite{10.1093/mnras/stw483} \\
J1721-2457 &    0.387 &   6.751 &  1.393 &  12.7 &      -30(70) &           0.01  & \cite{10.1093/mnras/stw483} \\
J1730-2304 &    3.137 &   6.023 &  0.620 &  16.9 &       0(2) &            0  & \cite{10.1093/mnras/stv2395} \\
J1732-5049 &  340.029 &  -9.454 &  1.873 &   8.0 &         20(20) &        0.03  &    \cite{10.1093/mnras/stv2395} \\
J1738+0333 &   27.721 &  17.742 &  1.471 &   7.3 &      -30(90) &          0  & \cite{10.1093/mnras/stw483} \\
J1744-1134 &   14.794 &   9.180 &  0.395 &  17.3 &         0.8(8) &          0.02  & \cite{10.1093/mnras/stw483} \\
J1751-2857 &    0.646 &  -1.124 &  1.087 &   8.3 &      -10(50) &           0 & \cite{10.1093/mnras/stw483} \\
J1801-1417 &   14.546 &   4.162 &  1.105 &   7.1 &  -30(100) &           0.02 & \cite{10.1093/mnras/stw483} \\
J1802-2124 &    8.382 &   0.611 &  0.760 &   7.2 &       10(60) &           0.01 & \cite{10.1093/mnras/stw483} \\
J1804-2717 &    3.505 &  -2.736 &  0.805 &   8.1 &      -40(40) &           0 & \cite{10.1093/mnras/stw483} \\
J1843-1113 &   22.055 &  -3.397 &  1.260 &  10.1 &         -7(20) &        0.05  &   \cite{10.1093/mnras/stw483} \\
J1853+1303 &   44.875 &   5.367 &  2.083 &   8.4 &        -30(20) &         0  &  \cite{10.1093/mnras/stw483} \\
B1855+09   &   42.290 &   3.060 &  1.200 &  17.3 &       1(2) &           0.03  & \cite{10.1093/mnras/stw483} \\
J1909-3744 &  359.731 & -19.596 &  1.140 &   9.4 &         0.6(9) &     0.02     &   \cite{10.1093/mnras/stw483} \\
J1910+1256 &   46.564 &   1.795 &  1.496 &   8.5 &         30(20) &          0  & \cite{10.1093/mnras/stw483} \\
J1911+1347 &   25.137 &  -9.579 &  1.069 &   7.5 &          14(8) &           0  & \cite{10.1093/mnras/stw483} \\
J1911-1114 &   47.518 &   1.809 &  1.365 &   8.8 &       20(50) &          0   & \cite{10.1093/mnras/stw483} \\
J1918-0642 &   30.027 &  -9.123 &  1.111 &  12.8 &           0(8) &        2.46     & \cite{10.1093/mnras/stw483} \\
B1953+29   &   65.839 &   0.443 &  6.304 &   8.1 &      -20(50) &       0      & \cite{10.1093/mnras/stw483} \\
J2010-1323 &   29.446 & -23.540 &  2.439 &   7.4 &         20(20) &        0    & \cite{10.1093/mnras/stw483} \\
J2019+2425 &   64.746 &  -6.624 &  1.163 &   9.1 &      -500(900) &      0     &  \cite{10.1093/mnras/stw483} \\
J2033+1734 &   60.857 & -13.154 &  1.740 &   7.9 &  40(100) &         0    & \cite{10.1093/mnras/stw483} \\
J2124-3358 &   10.925 & -45.438 &  0.410 &  16.8 &       0(3) &         0.02     & \cite{10.1093/mnras/stv2395} \\
J2129-5721 &  338.005 & -43.570 &  3.200 &  15.4 &      -1(2) &         0    &  \cite{10.1093/mnras/stv2395} \\
J2145-0750 &   47.777 & -42.084 &  0.714 &  17.5 &      -2(1) &        0.28    & \cite{10.1093/mnras/stw483} \\
J2229+2643 &   87.693 & -26.284 &  1.800 &   8.2 &        -20(20) &    0    &     \cite{10.1093/mnras/stw483} \\
J2317+1439 &   91.361 & -42.360 &  1.667 &  17.3 &      -1(3) &          0   & \cite{10.1093/mnras/stw483} \\
J2322+2057 &   96.515 & -37.310 &  1.011 &   7.9 &       30(70) &        0    & \cite{10.1093/mnras/stw483} \\

\bottomrule
\end{tabular}
\end{center}
\end{table}

\end{document}